\documentclass{aastex701}

\usepackage{hyperref, graphicx, amsmath}
\usepackage{xcolor}

\begin{document}


\title{Validating the Angular Sizes of Red Clump Stars with Intensity Interferometry}

\correspondingauthor{Alex G.\ Kim}

\author[0000-0001-6315-8743]{Alex G.\ Kim}
\email{agkim@lbl.gov}
\affiliation{Physics Division, Lawrence Berkeley National Laboratory\\
1 Cyclotron Road, Berkeley, CA 94720 USA}
\author[0000-0001-5194-3680]{Robin Kaiser}
\email{robin.kaiser@univ-cotedazur.fr}

\affiliation{Universit{\'e} C{\^ o}te d’Azur, CNRS, Institut de Physique de Nice, 06200 Nice, France}

\begin{abstract}
The surface-brightness-color (SBC) relationship for Red Clump stars provides a critical foundation for precision distance ladder measurements, including the 1\% distance determination to the Large Magellanic Cloud. Current SBC calibrations rely on angular diameter measurements of nearby Red Clump stars obtained through long-baseline optical interferometry using the Very Large Telescope Interferometer. We explore the application of intensity interferometry to measure limb-darkened angular diameters of Red Clump stars, offering a complementary approach to traditional amplitude interferometry. We describe the framework for extracting angular diameters from squared visibility measurements in intensity interferometry, accounting for limb darkening through the stellar atmosphere models. For the Red Clump star HD~17652, we show that intensity interferometry in the $H$ band at baselines matching PIONIER ($\sim$100~m) could achieve $<1$\% angular size uncertainties in 2-hour exposures by measuring the primary peak of the visibility function, enabling direct comparison with existing measurements. Critically, observations at shorter wavelengths probe the secondary visibility maximum, providing independent checks of both measurement and systematic errors that are largely insensitive to limb-darkening assumptions. Exploiting the multiplex advantage of simultaneous multi-bandpass observations and the large number of baselines available with telescope arrays such as the Cherenkov Telescope Array Observatory can reduce observing times to practical levels, making intensity interferometry a viable tool for validating the angular sizes for a subset of the Red Clump star calibration sample.
\end{abstract}



\section{Introduction}

Red Clump stars represent an evolutionary phase for intermediate-mass stars 
that have exhausted their core hydrogen and evolved beyond the red giant branch to burn helium stably in their cores as horizontal branch objects. These stars occupy a distinctive region in the Hertzsprung-Russell diagram, forming a prominent clump of stars at effective temperatures around 5,000 K and luminosities of approximately 75~$L_\odot$, located on the cool side of the horizontal branch. This concentration creates the characteristic "Red Clump" feature that gives these stars their name.
Spectroscopically, Red Clump stars are classified as late G-type to early K-type giants, distinguished by their intermediate surface gravity ($\log g \approx 2.5-3.0$) and relatively high metallicity compared to globular cluster horizontal branch stars. They can be identified through their spectral characteristics, including the strength of pressure-sensitive lines such as Ca~I $\lambda$6162~\AA\  
and Sr~II $\lambda$4077~\AA\, which are diagnostic of their giant luminosity class and intermediate surface gravity.

Red Clump stars exhibit remarkably uniform intrinsic properties, particularly their surface-brightness-color (SBC) relationship \citep{1969MNRAS.144..297W, 2019Natur.567..200P}.
\citet{2019Natur.567..200P} calibrated the SBC relationship using angular diameters of a set of 48 stars. These stars have parallax distances measured by the Gaia satellite \citep{2016A&A...595A...1G} and are sufficiently nearby to resolve their angular diameters using long-baseline optical interferometry. 
The $S_V$ surface brightness is calibrated through
$V-K$ color with a 0.018~mag/arcsec$^2$ r.m.s.\ scatter.
Given this relationship, the angular size of a star can be deduced from its apparent magnitude and color.

For eclipsing binary systems containing Red Clump stars, the physical parameters of the stellar components -- including the semimajor axis, fractional radii, and absolute physical radius of each star -- can be derived from simultaneous fits to radial velocity curves obtained from high-resolution spectroscopy and photometric light curves \citep{2012ApJ...750..144G}.
The physical parameters of 20 detached eclipsing binaries (DEBs) in the Large Magellanic Cloud have been measured using this procedure \citep{2018ApJ...860....1G}. 
The ratio between the measured physical radius (from eclipse geometry) and the angular radius (derived from the SBC relationship) directly yields a 1\% precision distance to the LMC \citep{2019Natur.567..200P}.

\begin{figure}
\centering
\includegraphics[width=6.5in]{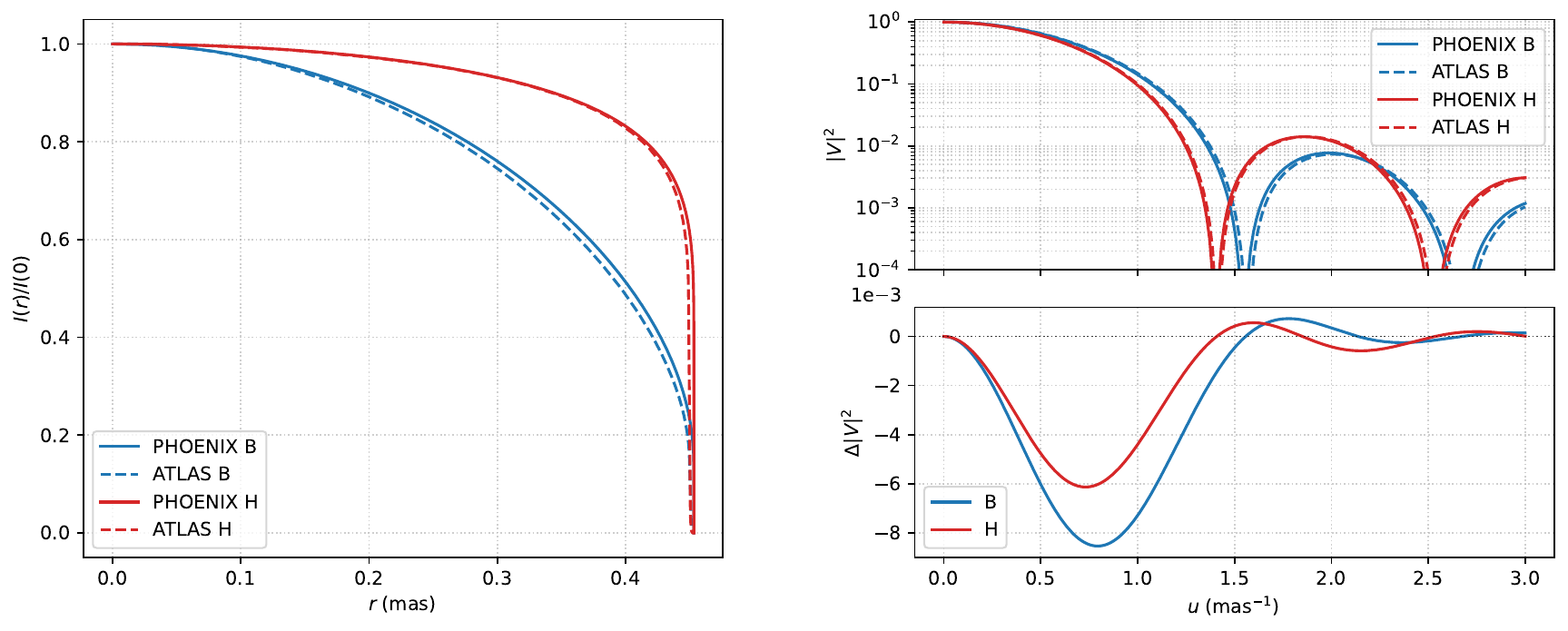}%
\caption{Left: Normalized $B$ and $H$-band intensity profiles $I(\theta)/I(0)$ computed by SATLAS and PHOENIX for $T=4800$K and $\log{g}=2.5$, which is the closest match to the Red Clump star HD~360.  Right top:  Squared visibility $|V|^2$ corresponding to the intensity profiles.
Right bottom: Difference between the PHOENIX and SATLAS squared visibilities.
The interferometric measurements of  HD~360 by
\citet{2018A&A...616A..68G} 
lie within $u<0.5$.
\label{fig:Intensity}}
\end{figure}

The SBC calibration is based on the angular sizes of calibration stars
as determined by \citet{2018A&A...616A..68G} 
through
measurements of squared visibility using the Very Large Telescope Interferometer (VLTI) \citep{2010SPIE.7734E..04H} with the four-telescope beam combiner PIONIER \citep{2011A&A...535A..67L}. 
The angular diameters of these stars span approximately 1--2~mas, with $V$-band magnitudes ranging from 4--6.5.
The observations utilized projected baselines ranging from 45 to 140~m across several spectral channels in the $H$ band (1.65 $\mu$m). Limb-darkened angular diameters were derived by fitting theoretical radial intensity profiles $I(\theta)$ from SATLAS stellar atmosphere models \citep{2013A&A...554A..98N}  to the observed squared visibility data \footnote{The model was downloaded using the Jean-Marie Mariotti Center (JMMC) — MOIO AMHRA service at \url{https://amhra.jmmc.fr/}.}.
As an example, the intensity profile and associated squared visibility
$|V|^2$ for the calculated grid point closest to the Red Clump calibrator HD~360, $T=4800$ and $\log{g}=2.5$, are shown in Figure~\ref{fig:Intensity}.

To resolve a 1~mas source at optical to near-infrared wavelengths requires a long interferometric baseline
(or an unfeasibly large telescope).  The Rayleigh resolution criterion is
defined  as the angular separation corresponding to the first zero of the diffraction pattern for a uniform disk source
\begin{equation}
    \theta = 1.22\frac{\lambda}{D},
\end{equation}
where $\lambda$ is the wavelength and $D$ is the baseline.  PIONIER with $H$-band central wavelength 1.65~$\mu$m, and a baseline of  $D=140$~m, can resolve $\theta = 2.97$~mas, several times larger than the 
angular diameters of the SBC calibrator stars.  Although marginally resolved according to the Rayleigh criterion, the stars' angular sizes can be precisely
determined by fitting
the shape of the theoretical intensity profiles predicted by the stellar atmosphere. 
As seen in
Figure~3 of
\citet{2018A&A...616A..68G} the interferometric measurements of $|V|^2$ of HD~360
lie in the central peak of the visibility function
(within a spacial frequency of $u<0.5$ in Figure~\ref{fig:Intensity}).
Nevertheless, the six baselines spanning 45-140~m obtained from the PIONIER telescope
configurations are sensitive to the  size-dependent shape of the
square-visibility function.

Angular diameter measurements depend on the assumed model of the stellar intensity profile. SATLAS modifies the plane-parallel stellar atmosphere code ATLAS to treat spherically extended geometry. ATLAS calculates line blanketing using opacity distribution functions, which average the contributions from different atomic and molecular species. \citet{2018A&A...616A..68G} estimate the model error contribution to angular size measurements by comparing intensity profiles at adjacent grid points in temperature, surface gravity, and stellar mass.
PHOENIX is an alternative stellar atmosphere code that is inherently spherically symmetric and calculates opacities directly from a comprehensive library of 500,000 spectral lines \citep{2013A&A...553A...6H}. Previous comparisons between PHOENIX and non-spherical ATLAS models have shown small but wavelength-dependent differences. For the M4 giant star $\psi$~Phoenicis, \citet{2004A&A...413..711W} found differences of $\sim$1.5\% in the resulting diameter values in the $K$-band using VLTI data. For the F5 star Procyon, \citet{2005ApJ...633..424A} found no differences in the $K$-band, but by 1.1\% at visible wavelengths (500nm) using data from the Mark III interferometer.
For our analysis, the PHOENIX intensity profile and squared visibility for the best-matched grid points are plotted alongside those of SATLAS in Figure~\ref{fig:Intensity} \citep{2023A&A...674A..63C}. By visual inspection, the only discernible difference appears in the intensity at the outer edge of the stellar disk. The difference between the squared visibilities is $\lesssim 0.006$ across both the plotted range and the spatial frequency range where the HD~360 data are fit. These differences are significantly smaller than the measurement uncertainties, demonstrating consistency between SATLAS and PHOENIX models for angular diameter determination of HD~360.

\begin{figure}
\centering
\includegraphics[width=5.5in]{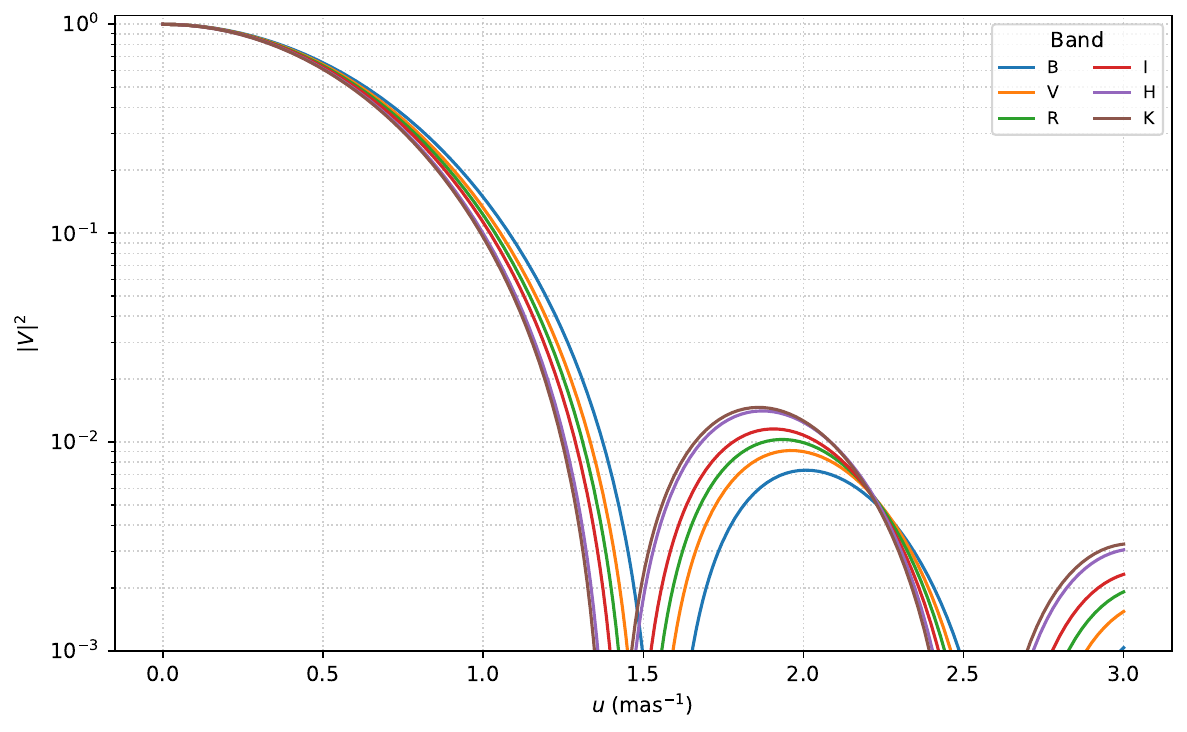}%
\caption{Squared visibilities $|\mathcal{V}|^2$
in $B$, $V$, $R$, $I$, $H$, and $K$ for the intensity profiles calculated by SATLAS. \label{fig:visibility_satlas}}
\end{figure}

Despite the apparent robustness of the angular size measurements from PIONIER, we suggest that there is  value in validating them with complementary interferometric measurements at multiple wavelengths that sample both the first zero and secondary maxima of the visibility function. For any given photometric band, the first zero and secondary maxima of the visibility curve provide a robust measure of the stellar angular diameter that are largely insensitive to model-dependent limb darkening effects, as demonstrated in the comparison between PHOENIX and ATLAS models shown in Figure~\ref{fig:Intensity}. The wavelength-dependence of the visibility amplitude, zero positions, and secondary maxima positions provides independent cross-checks of the angular size determination across the electromagnetic spectrum, as illustrated in Figure~\ref{fig:visibility_satlas}, which plots the squared visibility in $BVRIHK$ photometric bands for the intensity profiles calculated by SATLAS.
Such model-independent validation is relevant because the fitting accuracy of both ATLAS and PHOENIX stellar atmosphere models degrades significantly for G and K-type stars, including Red Clump stars, due to increased molecular opacity and convective effects that are difficult to model accurately \citep{2004AJ....128..829B}.
Reducing the model-dependence of the results through multi-wavelength visibility measurements would therefore substantially strengthen the reliability and precision of angular size measurements for these stellar types.

Systematic errors in $|V|^2$ measurements from amplitude interferometry, including PIONIER, present fundamental calibration challenges \citep{10.1093/mnras/stz114}. The primary systematic issue lies in determining the absolute normalization of visibility amplitudes, as PIONIER relies on calibrator stars to estimate instrumental transfer functions, but this process introduces systematic biases when calibrators have undetected binaries (potentially causing up to 2\% systematic errors in squared visibility) or when the transfer function varies between calibrator and science observations. Additional systematic sources identified at VLTI include differential polarization effects (several percent), atmospheric biases that affect spectral channels coherently, and data processing artifacts in Fourier space bias removal during PIONIER reduction pipelines. These systematic effects are particularly problematic for precision stellar diameter measurements because they impact all observations in a given sequence similarly and propagate through the calibration chain in highly correlated ways \citep{10.1093/mnras/stz114}. Recent high-precision diameter measurements have revealed systematic differences of several sigma and up to 15\% between well-resolved and marginally resolved interferometric observations, demonstrating that systematic errors can dominate the uncertainty budget for underresolved sources like Red Clump stars.

Intensity interferometry accommodates multi-frequency observations with long baselines that can resolve the visibility function of small astronomical sources. The Hanbury Brown--Twiss effect produces coherence in the photon arrival times from a given source, a coherence that depends on the source intensity profile, the telescope baseline geometry, and the observing frequency \citep{1956Natur.177...27B, 1956Natur.178.1046H}. The measurement depends on the timing of photons arriving at multiple locations, not their mixing, making long baselines more feasible than with traditional amplitude interferometry.
One of the most important science applications of intensity interferometry has been to measure the angular sizes of nearby stars, which helped establish the physical scale and fundamental properties of stellar objects \citep{1967MNRAS.137..393H, 1974MNRAS.167..121H}. These pioneering measurements provided the first direct determinations of stellar diameters and established the effective temperature scale for early-type stars, fundamentally advancing stellar astrophysics \citep{1976ApJ...203..417C}.
In the current era, kilometer-scale intensity interferometers utilizing arrays of optical air Cherenkov telescopes can achieve microarcsecond angular resolution, which will reveal unprecedented details across stellar surfaces and in their immediate circumstellar environments \citep{2012MNRAS.419..172N, 2016SPIE.9907E..0MD}.
Measurements of stellar diameters are coming from the
Cherenkov telescopes VERITAS,
MAGIC, and H.E.S.S.\ 
\cite{2020MNRAS.491.1540A, 2024MNRAS.52712243Z,  2024ApJ...966...28A, 2025MNRAS.537.2334V}
and deployments at 1m to 2m class telescopes
\cite{2021MNRAS.506.1585Z, 2023AJ....165..117M}.

An advantage of intensity interferometry over amplitude interferometry is its intrinsic independence from electromagnetic phase information. Unlike amplitude interferometry, which measures the coherent superposition of light waves and therefore requires exquisite phase stability, intensity interferometry measures second-order correlations in photon arrival times.
While atmospheric turbulence, mechanical vibrations, and optical path instabilities do affect arrival times, they are currently dwarfed by the dominant contribution of jitter in current detectors, and thus do not significantly contribute to the systematic error budget of intensity interferometry.
The technique does not require combining light coherently, eliminating systematic errors associated with beam combination optics, polarization mismatch, and the complex calibration chains needed to extract visibility amplitudes and closure phases. However, intensity interferometry introduces its own set of systematic challenges.
\added{Clock synchronization between telescopes, achievable to better than 100~ps using, e.g., White Rabbit electronics, and baseline determination, achievable to better than 1~cm with Real-Time Kinematic GPS systems, are no longer leading limitations.
The primary sources of systematic error in current IACT-based implementations are the following.
(1)~\textit{Poor point-spread function and large camera pixels}: IACT cameras are designed for air-shower imaging and have large pixels, causing the stellar signal to be measured simultaneously with sky background light and biasing the $g^{(2)}$ function.
In practice, the H.E.S.S.\ and MAGIC SII programs have mitigated this by mounting dedicated external reimaging optics at the IACT focal plane, bypassing the native camera \citep{2024MNRAS.529.4387A,2024MNRAS.52712243Z,2025MNRAS.537.2334V}.
(2)~\textit{Filter bandpass shift with incidence angle}: the fast focal ratios of IACT optics ($f/\sim 0.5$ for the CTAO SSTs, which use a Schwarzschild--Couder dual-mirror design) mean that photons from different parts of the primary mirror converge to the focal plane at angles up to $\sim 45^\circ$ from normal, shifting interference filter passbands by several percent across the aperture and increasing the background from out-of-band photons at the mirror edges; this limits the minimum usable bandwidth at optical wavelengths.
(3)~\textit{Finite mirror size}: when the first null of the visibility function falls within the baseline spanned by a single mirror, the measured zero-baseline $g^{(2)}$ is distorted; this is most significant for large ($\gtrsim 3$--4~mas) stars observed with 10~m-class IACT mirrors.
For our Red Clump targets ($\theta \approx 1$--2~mas), the first $H$-band null falls at $\sim 150$--300~m, so the visibility variation across the 4.3~m CTAO SST primary is $\lesssim 0.04\%$---negligible for our application.
(4)~\textit{Background noise and multi-night systematics}: accidental coincidences bias the correlation function for faint sources, and night-to-night variations in sky brightness can cause systematic drifts in the $g^{(2)}$ baseline when observations are combined across many sessions, degrading the S/N below the $1/\sqrt{T}$ expectation.}

The impact of systematic errors on squared visibility measurements varies  depending on the baseline regime. As evident from Figure 2, measurements obtained along the steeply declining slopes of the visibility function are highly sensitive to systematics that affect the amplitude, since a systematic change in the amplitude or slope of $|\mathcal{V}^2|$ translates to a change in the baseline-dependent slope upon which the stellar diameter determination depends. Conversely, measurements obtained near the local maxima of  $|\mathcal{V}^2|$—such as at zero baseline or at baselines that
cover the secondary peaks arising from the limb-darkened stellar disk—exhibit substantially reduced sensitivity to distortions caused by systematics. At these critical points, the gradient $\partial |\mathcal{V}^2|/\partial u$ approaches zero, making the measured baseline position of the extrema less sensitive to calibration errors. This behavior is analogous to measurements of the Baryon Acoustic Oscillation (BAO) peak in the cosmic matter power spectrum, where the characteristic scale of the acoustic feature can be robustly determined even in the presence of errors that can affect the overall shape. This suggests an observational strategy where multiple baselines sampling the local extrema of the visibility function could provide internally consistent diameter estimates that are less susceptible to systematic contamination than measurements relying solely on the steep visibility slope.

With interferometry in optical bands as opposed to $H$, the baseline required to meet
the resolution criterion decreases.   For example, $D=138$~m is
required to resolve 1~mas in the $V$-band.
Several current and planned intensity interferometry facilities can achieve the baseline distances required for intensity interferometry of Red Clump stars. The Cherenkov Telescope Array Observatory (CTAO) represents a promising platform, with its planned arrays offering multiple telescopes separated by distances of 100-300 meters ideal for microarcsecond angular resolution \citep{2024SPIE13095E..27C}. The CTAO's Southern Array will include 37 4.3~m Small-Sized Telescopes,  improving measurement precision through simultaneous multi-baseline observations \citep{2024MNRAS.529.4387A}. 
Discussions are underway to connect the Very Large Telescope's Unit Telescopes and the VISTA telescope.
Additional promising sites for intensity interferometry include using additional baselines at Calern
Plateau \citep{2022SPIE12183E..0GM}, linking existing telescopes
at Mauna Kea \citep{presentation_kaiser},
Cerro Pach\'{o}n, Cerro Tololo, and Las Campanas observatories in Chile, and the  Gran Telescopio Canarias and William Herschel Telescope telescopes at La Palma.
Additionally, existing amplitude interferometry facilities that could operate at optical wavelengths with such baselines include VLTI and the Center for High Angular Resolution Astronomy (CHARA) array at Mount Wilson Observatory \citep{2005ApJ...628..453T,2020SPIE11446E..05S} [Robin].

In Section II we present the theoretical framework for extracting limb-darkened angular diameters from intensity interferometry measurements, developing the mathematical formulation that relates stellar intensity profiles to squared visibility functions and deriving the Fisher information bounds for angular diameter precision. In Section III we demonstrate the application of this framework to validate surface-brightness-color relationship determinations for Red Clump stars, showing that intensity interferometry can achieve angular-diameter uncertainties of 3\% for representative targets like HD~17652 using realistic instrumental parameters, and that multi-wavelength observations sampling visibility nulls and secondary maxima provide model-independent cross-checks of the angular diameter measurements that underpin precision distance scale calibration.

\section{Intensity Interferometry Measurement of Angular Diameters}
Interferometry involves measuring a source during an exposure time with a telescope at position $\mathbf{x}$ separated from another telescope by a baseline $\mathbf{B}$.
The source is conveniently described by its normalized first- and second-order coherence functions
\begin{align}
    g^{(1)}(\mathbf{B},\Delta t) & = \frac{\langle E^*(\mathbf{x},t) E(\mathbf{x}+\mathbf{B}, t+\Delta t)\rangle}
    {\sqrt{\langle I(\mathbf{x},t)\rangle \langle I(\mathbf{x}+\mathbf{B}, t+\Delta t)\rangle}}\\
    g^{(2)}(\mathbf{B},\Delta t) & = \frac{\langle E^*(\mathbf{x},t)  E^*(\mathbf{x}+\mathbf{B}, t+\Delta t) E(\mathbf{x}+\mathbf{B}, t+\Delta t) E(\mathbf{x},t) \rangle}
    {\langle I(\mathbf{x},t)\rangle \langle I(\mathbf{x}+\mathbf{B}, t+\Delta t)\rangle}.
\end{align}
$g^{(1)}$ describes the correlation between the electric field (for a single polarization) at two telescopes, relevant for amplitude interferometry,
whereas $g^{(2)}$ describes the correlation between photon detection, relevant for intensity interferometry.

For chaotic light sources, such as stars, the coherence functions
depend only on the intensity.  When considering a narrow bandwidth $\Delta \nu$, over which
the intensity can be treated as being constant, the spatial and temporal components are separable.
For an axially symmetric source with
intensity $I(\theta)$, where $\theta$ is the angle from the star center,
\begin{align}
    g^{(1)}(u) & = \mathcal{V}(u) = \frac{ \int_0^\infty{I(\theta) J_0(2 \pi u\theta) \theta d\theta}}
    {\int_0^\infty{I(\theta)\theta d\theta}} \\
    g^{(2)}(u, \Delta t) & = 1 + |\mathcal{V}(u)|^2 |g^{(1)}(\Delta t)|^2, \label{eq:g2}
\end{align}
where $u=|\mathbf{B}|/\lambda$ for wavelength
$\lambda$
and $J_0$ is the Bessel function of the first kind.
$|g^{(1)}(\Delta t)|^2 = \text{sinc}^2{(\pi\Delta\nu \Delta t)}$, the Fourier transform
of the unit top-hat function squared.
The first equation is from the van Cittert-Zernike theorem and the second is the Siegert relation.  $\mathcal{V}(u)$ is often referred to as the normalized visibility.

The measurement in intensity interferometry is the correlation between photon arrival times at two detector positions.  It is generated \added[my]{by} making the normalized histogram of the difference in photon arrival times at the two telescopes.  The expected signal for
the source is given by $g^{(2)}$ in Eq.~\ref{eq:g2}.   The $|g^{(1)}(\Delta t)|^2$ term contributes to the height and shape of the histogram, while $|\mathcal{V}(u)|^2$ contributes to the height. 
The measurement is only possible because of the Hanbury Brown and
Twiss effect, i.e., $|g^{(1)}(\Delta t)|^2 \ne 0$.
The intensity profile information is encoded in the visibility.
The squared visibilities $|\mathcal{V}|^2(u) $ predicted for representative Red Clump stars are shown in Figures~\ref{fig:Intensity} and \ref{fig:visibility_satlas}.

The expected signal to noise ratio can be expressed as
\begin{equation}
    \text{SNR} = \frac{|\mathcal{V}|^2}{\sigma_{|\mathcal{V}|^2}},
\end{equation}
where 
\begin{equation}
\sigma^{-1}_{|\mathcal{V}|^2} =\frac{d\Gamma}{d\nu}
\left(
\frac{T_{\text{obs}}}{\sigma_t}
\right)^{1/2} 
\left(128 \pi\right)^{-1/4}, \label{eq:noise}
\end{equation}
$\Gamma$ is the mean rate of photons, $d\Gamma/d\nu= \epsilon AF_\nu/(h\nu_0)$, where $\epsilon$ is the instrumental throughput,
$A$ the collecting area of the telescope, and $F_\nu$ the specific flux.  The observing time is $T_{\text{obs}}$ and $\sigma_t$
is the detector timing jitter.
This expression for the signal to noise ratio is valid under the following conditions: the source is chaotic and unpolarized, the observational bandpass (centered at frequency $\nu_0$ with bandwidth $\Delta \nu$) is sufficiently narrow that the visibility does not vary significantly across the bandpass, and yet sufficiently broad to satisfy $\sigma_t \Delta \omega \gg 1$, where $\Delta \omega = 2\pi\Delta\nu$
\citep{2024PhRvD.109l3029D}.
This latter condition means that the detector timing jitter blurs
the timing more than the different frequencies transmitted through the finite bandwidth.

In the opposite regime where detector jitter is negligible, $\sigma_t \Delta \omega \ll 1$, 
\begin{equation}
\sigma^{-1}_{|\mathcal{V}|^2} = \frac{d\Gamma}{d\nu} \left(T_{\text{obs}} \Delta \omega\right)^{1/2}  \left(12 \pi \right)^{-1/2}.
\end{equation}
This noise depends on bandwidth $\Delta \omega$ and is lower than that of Equation~\ref{eq:noise}. However, achieving this regime would require spectral resolution $R>10000$ to get narrow enough passbands for the timing jitters of $>10$~ps characteristic of current detectors to be subdominant.  Given the complications of building a high-dispersion instrument, we restrict ourselves to the detector-jitter-dominated case.

We evaluate and present in Table~\ref{tab:inverse_noise} inverse noises for observations with two
4-m telescopes in several passbands.
The instrumental parameters are: a photon-counting detector with 42.4~ps FWHM timing jitter
\cite{10.1117/1.JATIS.11.3.035005} (corresponding to $\sigma_t=16.6$~ps standard deviation),
an overall throughput of 0.3, and an integration time of $T_{obs}=\added[1]{2}$~h.
The $\sigma_t \Delta \omega \gg 1$ condition translates to a condition on the bandwidth,
$\Delta \nu \gg 9.6$~GHz in frequency or $\Delta \lambda \gg 0.32$\AA\ at 1 micron wavelength.
Two stars, HD~360 and HD~17652, are presented as representative examples spanning the faint/small and bright/large extremes of the sample. 
Specific fluxes in the $V$, $H$, and $K$ bands for the target stars are taken from \citet{2018A&A...616A..68G, 2019Natur.567..200P}. Fluxes in the $R$ and $I$ bands are interpolated in wavelength space from the measured $V$, $H$, and $K$ fluxes together with Gaia $G$, $BP$, and $RP$ magnitudes.  These magnitudes are given 
in Table~\ref{tab:extended_data_2}.

\begin{table}
\centering
\begin{tabular}{lccccc}
\hline
Star & $V$ & $R$ & $I$ & $H$ & $K$ \\
\hline
HD 360 & 7.07 & 9.81 & 17.02 & 47.31 & 41.82 \\
HD 17652 & 28.93 & 40.37 & 69.70 & 188.53 & 168.63 \\
\hline
\end{tabular}
\caption{Inverse noise $\sigma^{-1}_{|\mathcal{V}|^2}$  for the stars HD~360 and HD~17652 for two
4-m telescopes in $VRIHK$ passbands.
The instrumental parameters are: a photon-counting detector with 42.4~ps FWHM timing jitter,
an overall throughput of 0.3, and an integration time of $T_{obs}=2$~h.  This is equivalent to the signal-to-noise ratio at zero baseline. }\label{tab:inverse_noise}
\end{table}

The parameter uncertainty of $s$ is bounded using the Fisher information matrix through the Cram\'er-Rao bound
\begin{equation}
\text{Var}(s) \ge F_{ss}^{-1} =   \left(\sum \left(2 \mathcal{V}\frac{\partial \mathcal{V}}{\partial s}  \right)^2 \sigma_{|\mathcal{V}|^2}^{-2} \right)^{-1},
\end{equation}
where the sum is over all measurements and
\begin{equation}
    \frac{\partial \mathcal{V}}{\partial s} = -\frac{2\pi u}{s}
     \frac{ \int_0^\infty{I(\theta) J_1(2 \pi u\theta) \theta^2 d\theta}}
    {\int_0^\infty{I(\theta)\theta d\theta}}.
\end{equation}

The intensity profile of a star is given by $I(\theta)$, where
$\theta$ represents the angular separation between the stellar center and a given point on the stellar disk as seen from the observer and axial symmetry is assumed.
The SATLAS
library provides a model for the normalized profile as $I_0(\theta)$ for
a model red giant stars with angular diameter
$\theta_0$, on a grid of effective temperature, surface gravity, and mass.
Making the angular diameter a free parameter,
the model profile can be expressed in terms of the SATLAS model as
\begin{equation}
    I(\theta; s) \propto I_0(\theta/s),
\end{equation}
for a star with angular diameter $s\theta_0$.  The
parameter $s$ represents the angular scale of the star.

The uncertainty in the fit parameter $s$, the relative size of the star, is estimated 
as $\sigma_s = \sqrt{F_{ss}^{-1}}$, representing the Cram\'er-Rao lower bound.
The expected measurement precision for the observation configuration described above
and the two stars, HD~360 and HD~17652, are presented as representative examples spanning the faint/small and bright/large extremes of the sample. HD~360 has an angular diameter $\theta = 0.906$~mas and $V = 5.986$~mag, while HD~17652 has $\theta = 1.835$~mas and $V = 4.456$~mag. The stars have effective temperatures of 4764K and 4786K, respectively.

\begin{figure}
\centering
\includegraphics[width=5.5in]{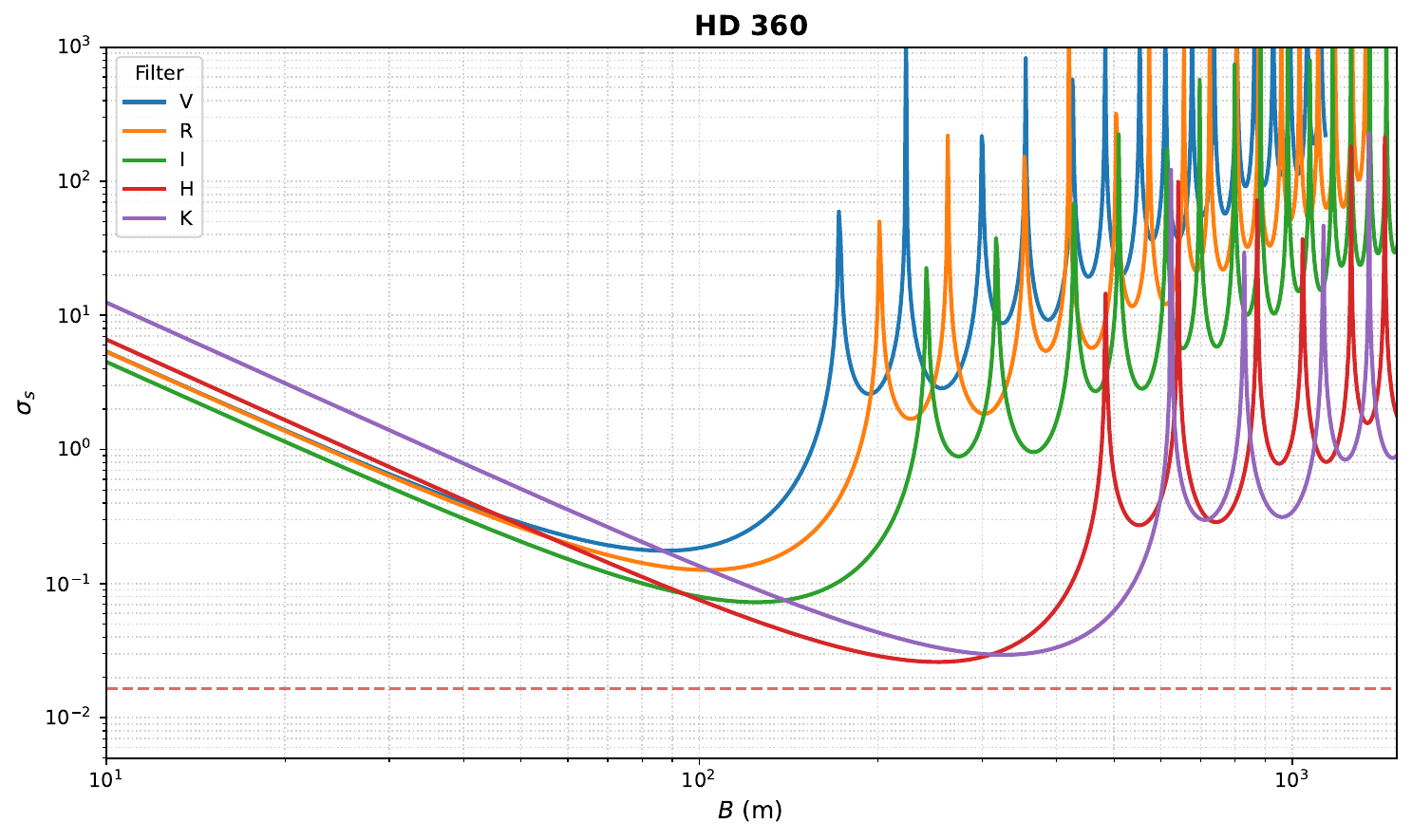}\\
\includegraphics[width=5.5in]{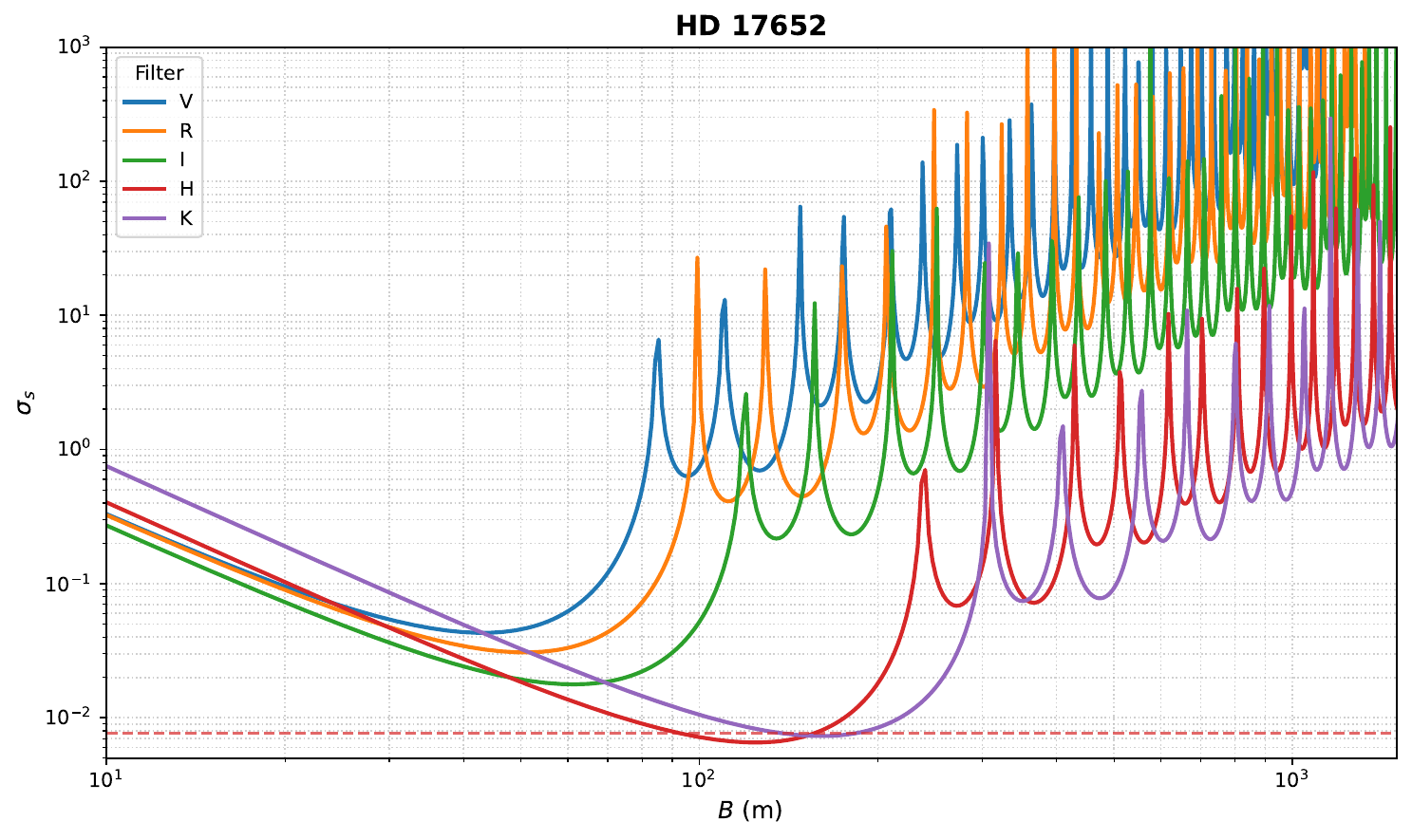}%
\caption{Estimated uncertainty in the scale factor parameter $s$. $\sigma_s$ for (Top) HD~360  and (Bottom) HD~17652 .  
The measurement noise is based on two
4-m telescopes separated by a baseline distance $B$, a photon detector detector
with 42.4~ps FWHM timing jitter,an overall instrumental
throughput of 0.3, and an exposure time is set to $T_{obs}=2$~h. 
The dashed horizontal line is the scale uncertainty for these stars as measured in the $H$-band using PIONIER
\citep{2018A&A...616A..68G}.
\label{fig:sigma_s}}
\end{figure}

Figure~\ref{fig:sigma_s} shows the scale parameter uncertainty $\sigma_s$ as a function of baseline $B$ for several photometric bands. The measurement precision is strongest in the $H$ band for both stars, as their cool temperatures produce peak flux in the near-infrared where photon counts are highest.
For HD~17652, the uncertainty in $s$ reaches a minimum of $\sigma_s \sim 0.1$ at baselines of $B \sim 50$--70~m in optical bands, before the star becomes resolved (the first null in the visibility function occurs) at $B \sim 100$~m. In the $H$ and $K$ bands, the measurement precision improves to $\sigma_s \lesssim 0.01$ at $B \sim 150$~m, with the first null occurring at $B \sim 300$~m. 
HD~360, being smaller and fainter, requires commensurately longer baselines to achieve comparable angular resolution and exhibits higher uncertainties due to lower photon flux.

\section{Validation of the SBC Determination}

The Red Clump surface-brightness-color relationship measured by \citet{2019Natur.567..200P} is based on high signal-to-noise data, results in small residual dispersion, and
demonstrates internal consistency.
The per-star angular diameter uncertainties have a mean of $0.015$~mas with a standard deviation of $0.006$~mas; the fractional (logarithmic) uncertainties, which correspond to uncertainties in our $s$ parameter, have a mean of $0.012$ and standard deviation of $0.006$.
The  uncertainties for the specific cases of HD~360 and HD~19652 are plotted in
Figure~\ref{fig:sigma_s}.
Note that fractional angular-diameter uncertainties transform directly to relative distance uncertainties.
The per-star surface brightness uncertainties have a mean of $0.03$~mag/arcsec$^2$ with a standard deviation of $0.01$~mag/arcsec$^2$. The SBC relationship derived from these measurements exhibits an r.m.s.\ scatter of $0.018$~mag/arcsec$^2$, comparable to the individual measurement uncertainties and implies a distance precision of 0.8\%. 
Furthermore, no extreme outlier stars deviate from the fitted SBC relationship.

However, the systematic errors mentioned earlier could potentially impact the normalization of the  $|\mathcal{V}|^2$ measurement  and the angular diameter fits, leading to an offset in the SBC determination.
A consistency check with intensity interferometric measurements, which have a distinct systematic error propagation into angular diameter, would provide a direct constraint on the systematic component of the SBC relation error budget.
Any potential bias between PIONIER and intensity interferometry angular diameters would be revealed by a statistically significant non-zero offset in the ensemble distribution of per-star size differences. To optimize the precision of the size-difference uncertainties, the PIONIER and intensity interferometry measurements should contribute equally to the total error; therefore, the new measurements should aim to achieve comparable per-star uncertainties to those of the PIONIER measurements.

Different intensity interferometry observations  provide varying tests of systematics.
The most direct comparison would use measurements that replicate those of PIONIER: in the $H$-band with baselines from 45--140~m.  Any inconsistencies in angular diameters would be due
to  error in the processing and interpretation of either data and be insensitive to stellar modeling.
As seen in Figure~\ref{fig:sigma_s}, 
for the relatively large and bright HD~17652, the lowest uncertainty in $s$ for a 2-hour exposure time is $\sigma_s < 0.007$ and is slightly
lower than that of \citet{2018A&A...616A..68G} occurring in the $H$ band at a baseline of slightly over 100~m.   For the relatively small and faint HD~360,  the lowest uncertainty in $s$ is $\sigma_s < 0.03$ and is
slightly higher than that of \citet{2018A&A...616A..68G} at a baseline of slightly over 100~m. 

Limb-darkening models predict intensity profiles that are wavelength-dependent, as illustrated in  Figure~\ref{fig:Intensity}.   Observations in multiple wavelengths not only collectively contribute to the measurement of angular diameters but also provide a means to reduce systematic errors due to modeling or enable comparisons between models.
The uncertainties in $s$ in Figure~\ref{fig:sigma_s} relative to the $H$-band increases by a factor $\sim 2$--10
going blueward from $I$ to $V$ bands, with decreasing optimal baselines, whereas in the $K$-band  $\sigma_s$ slightly increases with an optimal baseline that is doubled.  To recover the minimal $H$-band $\sigma_s$ in other bands, which have
2--10 times larger uncertainties at the same exposure time, would require integration times that are
4--100 times longer than the fiducial.

With longer baseline observations, it becomes possible to approach the Rayleigh resolution criterion, which, in principle, could yield angular diameters that are less sensitive to limb-darkening modeling. However, the resolution criterion is technically achieved when the visibility reaches its first zero, at which point the signal vanishes. As a practical alternative, we consider measuring the secondary maximum, which, when combined with the primary-maximum measurement, provides a means to reduce systematic errors or enable comparisons between models at a single wavelength. 
Here, the uncertainties in $s$
in Figure~\ref{fig:sigma_s} relative to the per-band minimum are larger by a factor of $\sim 10$, meaning that
a 100-times longer exposure time would be required to recover the minimum uncertainty.

There are several approaches to extend the fiducial observation to increase the
effective exposure time by factors of 4--10000 to achieve the target signal to noises at wavelengths
other than the $H$ band or at the secondary peak of the visibility function.
The most straightforward is to dedicate the telescope time for the longer exposures.  Otherwise,
one could exploit the multiplex advantage of observing multiple bandpasses simultaneously---for example, the requisite signal could be obtained by increasing the effective exposure time by dispersing the light dispersed onto 4 to 100 channel SPAD arrays (ensuring that the $\sigma_t \Delta \omega \gg 1$ condition holds)
and increasing the exposure time to make up for
what could not be gained through multiplexing.
Interferometry with simultaneous observation of $N$ telescopes offers ${N \choose 2}$ unique baselines. For $N=37$ telescopes \citep[the number of Small-Sized Telescopes planned for the Cherenkov Telescope Array Observatory-South][]{2022SPIE12182E..0KT}, this yields an effective increase by a factor of 666 to the nominal exposure time. \added{Crucially, simultaneous multi-baseline observations are preferable to extending single-pair integrations across multiple observing sessions: night-to-night variations in sky background can cause systematic drifts in the $g^{(2)}$ baseline that prevent the signal-to-noise ratio from scaling as $1/\sqrt{T}$ when data are combined over many nights \citep{2024MNRAS.529.4387A}. The 666 simultaneous CTAO baselines therefore offer a practical path to the required precision within a single session, avoiding this multi-night systematic entirely.}

Validation
of the \citet{2019Natur.567..200P} result does not require re-measurement of
all stars, but rather testing for inconsistency in measured angular sizes of a sufficient large subset of stars.
If the validation sample has the same per-star surface brightness uncertainty of $0.03$\,mag/arcsec$^2$, the uncertainty on the mean surface brightness of $N$ stars scales as $0.03/\sqrt{N}$\,mag/arcsec$^2$. To detect a systematic offset comparable to the intrinsic scatter of the SBC relationship ($0.018$\,mag/arcsec$^2$) at $3\sigma$ significance would require $(3 \times 0.03/0.018)^2 \approx 25$ stars. These estimates suggest that a validation sample of approximately half of the 48 Red Clump stars with comparable measurement precision would provide sufficient statistical power to test for significant inconsistencies with the \citet{2019Natur.567..200P} calibration.
The stars with the most easily measurable sizes, those that are brightest to first order, can be prioritized for observation; this approach is justified as there is currently no evidence of bias in the existing SBC relationship as a function of observed or intrinsic size, magnitude, distance, or color.

\section{Conclusions}

We have examined the feasibility of using intensity interferometry to measure limb-darkened angular diameters of Red Clump stars, a stellar population crucial for precision distance scale calibration. The technique offers several advantages as a complement to traditional amplitude interferometry:

First, measurements at the primary peak of the visibility function in the $H$ band with baselines of approximately 100~m can directly replicate the PIONIER configuration, achieving comparable precision and enabling direct validation of existing measurements. This provides an independent cross-check of the angular diameters that underpin current SBC relationship calibrations.

Second, and perhaps more importantly, multi-wavelength observations that probe the secondary visibility maximum provide powerful diagnostics for systematic uncertainties. The first visibility null and secondary peaks are largely insensitive to detailed limb-darkening prescriptions, as demonstrated by the close agreement between PHOENIX and SATLAS models at these positions. By measuring angular diameters at multiple wavelengths that sample different features of the visibility function, we can assess and mitigate model-dependent systematics that limit the accuracy of single-wavelength measurements.

Third, modern detector technology and telescope arrays offer practical paths to achieving the required signal-to-noise ratios. The multiplex advantage of dispersing light onto multi-channel SPAD arrays can reduce observing times by factors of $\sim$100, while large telescope arrays like the Cherenkov Telescope Array Observatory provide hundreds of simultaneous baselines. For HD~17652 observations at bluer wavelengths, combining these advantages reduces the required observing time from $\sim$ 100-1000 hours to order one hour of single-channel exposure time.

Fourth, the model-independent validation provided by multi-wavelength intensity interferometry becomes particularly valuable for G and K-type stars, including Red Clump stars, where stellar atmosphere models face significant challenges in accurately representing molecular opacity and convective effects. By reducing reliance on model-dependent corrections, intensity interferometry can strengthen the reliability of the Red Clump star sample used for SBC calibration.
Conversely, intensity interferometry data provide powerful diagnostics of stellar atmospheric structure \citep{2004A&A...413..711W}. Dispersing the light into narrow spectral passbands covering specific absorption lines enables the measurement of wavelength-dependent stellar radii, which can reveal the vertical stratification of different chemical species in the stellar atmosphere \citep{presentation_Aufdenberg}. These measurements directly probe the validity of atmospheric models and can constrain physical processes such as convection, chromospheric extension, and opacity sources.

The Red Clump stars in the \citet{2019Natur.567..200P} calibration sample are well-suited for intensity interferometry measurements with foreseeable instrumentation. 
These stars and their characteristics
are given in Table~\ref{tab:extended_data_2}.  Having been observed at the VLTI, the stars predominantly have southern declinations.
As demonstrated by the representative examples HD~360 and HD~17652, these stars are sufficiently bright ($V \sim 4$--6~mag) and subtend large enough angular diameters ($\theta \sim 0.9$--1.8~mas) to enable precise angular diameter measurements.
The Fisher information analysis indicates that scale parameter uncertainties of $\sigma_s \lesssim 0.1$
are achievable with modest baseline configurations ($B \sim 50$--150~m) and realistic observing times;
this means that intensity interferometry
can obtain the same precision as PIONIER per star in the $H$-band.  
Varying degrees of improvements to the fiducial observing setup, increased exposure times, wavelength multiplexing, multiple baselines, can enable these precisions.
There is no evidence for systematic bias in the measured stellar surface brightnesses as a function
of angular diameter or magnitude, so the larger--brighter stars can be preferentially targeted.

\begin{table}[htbp]
\centering
\begin{tabular}{lccccccccc}
\hline
\hline
Star & RA & Dec & LD & $V$ & $H$ & $K$ & $G$ & $G_{\rm BP}$ & $G_{\rm RP}$ \\
 & (deg) & (deg) & (mas) & (mag) & (mag) & (mag) & (mag) & (mag) & (mag) \\
\hline
HD 360 & $2.072898$ & $-8.824267$ & $0.906 \pm 0.014$ & $5.986 \pm 0.005$ & $3.757$ & $3.653$ & $5.7161$ & $6.2276$ & $5.0448$ \\
HD 3750 & $9.966729$ & $-44.796215$ & $1.003 \pm 0.019$ & $6.004 \pm 0.005$ & $3.612$ & $3.485$ & $5.6986$ & $6.2520$ & $4.9956$ \\
HD 4211 & $11.051806$ & $-38.421153$ & $1.100 \pm 0.009$ & $5.877 \pm 0.028$ & $3.426$ & $3.295$ & $5.5776$ & $6.1534$ & $4.8595$ \\
HD 5722 & $14.682634$ & $-11.380039$ & $0.995 \pm 0.018$ & $5.618 \pm 0.012$ & $3.496$ & $3.381$ & $5.3585$ & $5.8446$ & $4.7112$ \\
HD 8651 & $21.170044$ & $-41.492704$ & $1.228 \pm 0.011$ & $5.418 \pm 0.006$ & $3.142$ & $3.019$ & $5.1293$ & $5.6539$ & $4.4506$ \\
HD 9362 & $22.813872$ & $-49.072018$ & $2.301 \pm 0.017$ & $3.943 \pm 0.006$ & $1.748$ & $1.638$ & $3.6709$ & $4.2119$ & $3.0016$ \\
HD 10142 & $24.614439$ & $-36.528792$ & $0.964 \pm 0.004$ & $5.938 \pm 0.009$ & $3.679$ & $3.557$ & $5.6592$ & $6.1767$ & $4.9770$ \\
HD 11977 & $28.734768$ & $-67.646977$ & $1.528 \pm 0.010$ & $4.686 \pm 0.011$ & $2.594$ & $2.486$ & $4.4194$ & $4.9082$ & $3.7848$ \\
HD 12438 & $30.310783$ & $-30.002314$ & $1.091 \pm 0.015$ & $5.344 \pm 0.007$ & $3.281$ & $3.176$ & $5.0951$ & $5.5600$ & $4.4722$ \\
HD 13468 & $32.899263$ & $-1.825569$ & $0.886 \pm 0.009$ & $5.934 \pm 0.013$ & $3.783$ & $3.666$ & $5.6801$ & $6.1648$ & $5.0257$ \\
HD 15220 & $36.647143$ & $-20.042165$ & $1.185 \pm 0.015$ & $5.881 \pm 0.006$ & $3.342$ & $3.199$ & $5.5369$ & $6.1423$ & $4.7953$ \\
HD 15248 & $35.716699$ & $-73.645740$ & $0.949 \pm 0.018$ & $6.001 \pm 0.003$ & $3.669$ & $3.553$ & $5.7000$ & $6.2386$ & $5.0082$ \\
HD 15779 & $38.039157$ & $-1.035031$ & $1.185 \pm 0.013$ & $5.357 \pm 0.013$ & $3.186$ & $3.067$ & $5.0935$ & $5.5955$ & $4.4389$ \\
HD 16815 & $40.167597$ & $-39.855498$ & $2.248 \pm 0.009$ & $4.109 \pm 0.003$ & $1.820$ & $1.706$ & $3.8217$ & $4.3671$ & $3.1358$ \\
HD 17652 & $42.273032$ & $-32.405189$ & $1.835 \pm 0.010$ & $4.456 \pm 0.005$ & $2.256$ & $2.139$ & $4.1799$ & $4.6882$ & $3.5129$ \\
HD 17824 & $42.759445$ & $-21.004108$ & $1.391 \pm 0.013$ & $4.764 \pm 0.010$ & $2.781$ & $2.668$ & $4.5227$ & $4.9889$ & $3.9090$ \\
HD 18784 & $45.292228$ & $-7.663325$ & $1.036 \pm 0.013$ & $5.748 \pm 0.004$ & $3.456$ & $3.353$ & $5.4710$ & $5.9946$ & $4.7948$ \\
HD 23319 & $55.708052$ & $-37.313829$ & $2.033 \pm 0.010$ & $4.583 \pm 0.005$ & $2.141$ & $1.995$ & $4.2501$ & $4.8389$ & $3.5322$ \\
HD 23526 & $56.539012$ & $6.803286$ & $0.915 \pm 0.020$ & $5.909 \pm 0.008$ & $3.744$ & $3.634$ & $5.6516$ & $6.1449$ & $4.9928$ \\
HD 23940 & $56.983635$ & $-30.168911$ & $1.093 \pm 0.020$ & $5.543 \pm 0.007$ & $3.344$ & $3.229$ & $5.2567$ & $5.7565$ & $4.5963$ \\
HD 30814 & $72.548613$ & $-16.216927$ & $1.310 \pm 0.008$ & $5.041 \pm 0.010$ & $2.898$ & $2.791$ & $4.7697$ & $5.2635$ & $4.1158$ \\
HD 36874 & $83.281322$ & $-35.139494$ & $1.118 \pm 0.010$ & $5.768 \pm 0.010$ & $3.364$ & $3.242$ & $5.4468$ & $6.0011$ & $4.7516$ \\
HD 39523 & $87.457574$ & $-56.166990$ & $1.939 \pm 0.013$ & $4.500 \pm 0.006$ & $2.160$ & $2.036$ & $4.1727$ & $4.7544$ & $3.4993$ \\
HD 39640 & $87.721778$ & $-52.109215$ & $1.251 \pm 0.016$ & $5.163 \pm 0.005$ & $3.040$ & $2.921$ & $4.8983$ & $5.3940$ & $4.2426$ \\
HD 39910 & $88.875836$ & $-4.616633$ & $1.090 \pm 0.006$ & $5.863 \pm 0.005$ & $3.451$ & $3.315$ & $5.5491$ & $6.1156$ & $4.8321$ \\
HD 40020 & $89.206515$ & $11.520808$ & $1.012 \pm 0.022$ & $5.876 \pm 0.003$ & $3.543$ & $3.419$ & $5.5960$ & $6.1384$ & $4.9029$ \\
HD 43899 & $94.255140$ & $-37.737073$ & $1.264 \pm 0.016$ & $5.557 \pm 0.015$ & $3.134$ & $3.004$ & $5.2169$ & $5.7830$ & $4.5128$ \\
HD 46116 & $96.369139$ & $-69.689408$ & $1.145 \pm 0.030$ & $5.373 \pm 0.005$ & $3.206$ & $3.103$ & $5.1133$ & $5.6067$ & $4.4649$ \\
HD 53629 & $106.196116$ & $-22.031844$ & $1.065 \pm 0.023$ & $6.085 \pm 0.006$ & $3.556$ & $3.410$ & $5.7438$ & $6.3377$ & $5.0087$ \\
HD 56160 & $108.712965$ & $-27.038125$ & $1.411 \pm 0.010$ & $5.580 \pm 0.000$ & $2.976$ & $2.823$ & $5.2113$ & $5.8384$ & $4.4564$ \\
HD 60060 & $112.498674$ & $-52.650917$ & $0.948 \pm 0.009$ & $5.872 \pm 0.002$ & $3.662$ & $3.545$ & $5.5971$ & $6.1015$ & $4.9329$ \\
HD 60341 & $113.331633$ & $-19.412838$ & $1.190 \pm 0.021$ & $5.645 \pm 0.015$ & $3.264$ & $3.126$ & $5.3287$ & $5.8823$ & $4.6301$ \\
HD 62713 & $115.925475$ & $-40.934506$ & $1.446 \pm 0.010$ & $5.134 \pm 0.017$ & $2.782$ & $2.654$ & $4.8145$ & $5.3657$ & $4.1204$ \\
HD 68312 & $122.887321$ & $-7.772651$ & $1.020 \pm 0.022$ & $5.359 \pm 0.008$ & $3.390$ & $3.279$ & $5.1193$ & $5.5721$ & $4.5099$ \\
HD 74622 & $130.586174$ & $-55.774374$ & $1.020 \pm 0.014$ & $6.279 \pm 0.010$ & $3.672$ & $3.532$ & $5.9088$ & $6.5086$ & $5.1706$ \\
HD 75916 & $133.128312$ & $-13.232945$ & $1.013 \pm 0.020$ & $6.117 \pm 0.009$ & $3.653$ & $3.516$ & $5.7979$ & $6.3678$ & $5.0819$ \\
HD 176704 & $285.615194$ & $-24.847612$ & $1.317 \pm 0.010$ & $5.645 \pm 0.005$ & $3.107$ & $2.956$ & $5.2825$ & $5.8831$ & $4.5515$ \\
HD 191584 & $303.099206$ & $-42.780502$ & $1.024 \pm 0.021$ & $6.211 \pm 0.012$ & $3.664$ & $3.512$ & $5.8598$ & $6.4595$ & $5.1213$ \\
HD 219784 & $349.706129$ & $-32.532377$ & $2.117 \pm 0.023$ & $4.412 \pm 0.008$ & $2.019$ & $1.886$ & $4.0769$ & $4.6631$ & $3.3808$ \\
HD 220572 & $351.331793$ & $-56.849071$ & $1.092 \pm 0.012$ & $5.605 \pm 0.020$ & $3.350$ & $3.224$ & $5.3153$ & $5.8379$ & $4.6387$ \\
HD 204381 & $322.181482$ & $-21.807204$ & $1.524 \pm 0.015$ & $4.501 \pm 0.008$ & $2.537$ & $2.426$ & $4.2685$ & $4.7183$ & $3.6513$ \\
\hline
\hline
\end{tabular}
\caption{Stars used to calibrate the Red Clump SBC relationship.  RAs and Decs are from Gaia \citep{2023A&A...674A...1G}.
The angular diameter LD and $V$ are
from \citet{2019Natur.567..200P}.  
The original source of the $H$ and $K$ magnitudes is \citet{2012MNRAS.419.1637L}.
The remaining columns are Gaia $G$-band magnitude, and Blue Photometer ($G_{\text{BP}}$) and 
    Red Photometer ($G_{\text{RP}}$)
    magnitudes.
\label{tab:extended_data_2}}

\end{table}

The local stars used to calibrate the SBC relationship establish an important rung of the cosmic distance ladder used to determine the Hubble constant $H_0$.
The more distant Large Magellanic Cloud (LMC) hosts Red Clump stars in detached eclipsing binary systems, whose physical radii are determined using spectroscopic and photometric measurements. The angular diameters of these LMC stars are inferred from the SBC relationship established by the calibrator stars. The distance to the LMC is then determined from the ratio of the angular and physical diameters \citep{2019Natur.567..200P}.
With this precisely known distance, the LMC serves as a crucial anchor for calibrating the Cepheid period-luminosity relationship, as it hosts thousands of Cepheid variable stars
\citep{Riess_2019}.  Similarly, the Tip of the Red Giant Branch has been absolutely calibrated
using stars in the LMC \citep{2018ApJ...858...12H}. These calibrated relationships then connect to subsequent distance rungs that extend to galaxies in the Hubble flow, enabling the determination of $H_0$.
Because SBC-based distances scale as $d \propto 1/s$, any systematic bias in the calibrators' fractional angular sizes, $s \rightarrow s(1+b_s)$, propagates with the opposite sign into the distance scale and therefore into the Hubble constant, $H_0 \rightarrow H_0(1-b_s)$ (i.e., $\delta H_0/H_0 \simeq -\,\delta s/s$).

Intensity interferometry also offers new measurements that can inform the modeling of binary systems. \citet{wqz5-2jjs} demonstrated that an extended-path intensity correlator—an optical modification that extends the traditional field of view of intensity interferometers—could determine all Keplerian orbital angles of a binary system to subdegree precision. When combined with radial velocity measurements, this technique can determine the masses of the constituent stars and the line-of-sight distance to per-mil precision or better. Intensity interferometry can thus be applied to test the robustness of current models of DEB systems.

Future work should focus on demonstrating these measurements with existing or planned facilities, particularly those offering both multi-wavelength capabilities and multiple baselines. Successful validation would establish intensity interferometry as a valuable tool for expanding the calibration sample of Red Clump stars with robust, cross-checked angular diameter measurements.

\begin{acknowledgments}
The authors acknowledge Marina Cortes, Andrew Liddle, Pierre Astier, Eric Linder, and Peter Nugent for their insightful discussions that contributed to this work.

A.~G.~Kim\ was supported by the U.S. Department of Energy (DOE), Office
of Science, Office of High Energy Physics, under Contract No. DE–AC02–05CH11231.
R.~Kaiser\ acknowledges the financial support the European project IC4Stars (ERC Advanced Grant No. 101140677).

This work has made use of data from the European Space Agency (ESA) mission
{\it Gaia} (\url{https://www.cosmos.esa.int/gaia}), processed by the {\it Gaia}
Data Processing and Analysis Consortium (DPAC,
\url{https://www.cosmos.esa.int/web/gaia/dpac/consortium}). Funding for the DPAC
has been provided by national institutions, in particular the institutions
participating in the {\it Gaia} Multilateral Agreement.
\end{acknowledgments}

\software{RedClump (v1.0; \url{https://github.com/AlexGKim/RedClump}), 
          direct\_plot.py}

\bibliography{main}
\bibliographystyle{aasjournalv7}

\end{document}